\documentclass[preprint,aps,showpacs]{revtex4}

\usepackage{graphicx}
\usepackage{bm}

\begin{document}

\title{ Statistical theory of spectra: Statistical moments as descriptors in
the theory of molecular similarity}

\author{Dorota Bieli\'nska-W\c{a}\.z}
\affiliation{Instytut Fizyki, Uniwersytet Miko\l{}aja Kopernika,
 Grudzi\c{a}dzka 5, 87-100 Toru\'n, Poland,
dsnake@phys.uni.torun.pl,
Fax: (++)(48-56)622-5397}

\author{ Piotr W\c{a}\.z }
\affiliation{ Centrum Astronomii, Uniwersytet Miko\l{}aja Kopernika,
 Gagarina 11, 87-100 Toru\'n, Poland}

\author{ Subhash C. Basak}
\affiliation{ Natural Resources Research Institute, 5013 Miller Trunk Highway,
 Minnesota 55811-1442 USA}

\date{\today}

\begin{abstract}
Statistical moments of the intensity distributions are used as molecular
descriptors. They are used as a basis for defining similarity distances
between two model spectra. Parameters which carry the information derived
from the comparison of shapes of the spectra and are related to the number
of properties taken into account, are defined.
\end{abstract}

\pacs{29.85.+c, 07.05.Kf, 33.20.-t, 33.70.-w}
\maketitle

keywords: molecular similarity, data analysis,
statistical theory of spectra, statistical moments

\section{Introduction}

The basic values in statistical theory of spectra are moments of the
intensity distribution ${\cal I}(E)$. In the case of discrete spectra the
$n-$th statistical moment is defined as:
\begin{equation}
M_n=\frac {\sum\limits_{i=1}^{max} {\cal I}_i(E) E_i^n}
{\sum\limits_{i=1}^{max} {\cal I}_i(E)},
\label{1}\end{equation}
where ${\cal I}_i$ is the intensity of the $i-$th line and $E_i$ is the
corresponding energy difference. If the spectral lines are sufficiently
close to each other then the spectrum may be approximated by a continuous
function. Then the $n-$th moment of the intensity distribution is defined as:
\begin{equation}
M_n=\frac{\int\limits_{C(E)}{\cal I}(E) E^n dE}
{\int\limits_{C(E)}{\cal I}(E) dE},
\label{2}\end{equation}
where $C(E)$ is the range of the energy for which the integrand
does not vanish. It is convenient to consider normalized spectra 
$I(E)=N {\cal I}(E)$, where $N=\left ( \int\limits_{C(E)}{\cal I}(E) dE
\right ) ^{-1}$,
for which
the area below the distribution function is equal to $1$.
Then
\begin{equation}
M_n=\int\limits_{C(E)}I(E) E^n dE.
\label{3}
\end{equation}
Convenient characteristics of the distributions
may be derived from the properly scaled distribution moments.
Moments normalized to the mean value equal to
zero ($M'_1=0$) are referred to as the {\em centered moments}.
The $n-th$ centered moment reads:
\begin{equation}
M^{\prime}_n=\int\limits_{C^{\prime}(E)} I(E) (E - M_1)^n dE.
\label{4}\end{equation}
The moments, for which additionally the variance is equal to $1$
$(M^{\prime\prime}_2=1)$ are defined
as
\begin{equation}
M^{\prime\prime}_n=\int\limits_{C^{\prime\prime}(E)} I(E)
\left [ \frac{(E - M_1)}{\sqrt{M_2 - M_1^2}} \right ]^n dE.
\label{5}\end{equation}

In this work the model spectra are approximated by continous functions
taken as linear combinations of $max$ unnormalized
Gaussian distributions centered at $\epsilon_i$
with dispersions $\sigma_i$, defined by the parameters
$c_i=1/2\sigma_i^2$, $i=1,2,\ldots max$:
\begin{equation}
I(E)=N \sum_{i=1}^{max} a_i exp \left[ -c_i(E - \epsilon_i)^2 \right] .
\label{6}\end{equation}
The normalization constant $N$ is determined so that
the zeroth moment of the distribution $I(E)$ is equal to $1$.

The $n$-th moment of the distribution is equal to:
\begin{equation}
M_n=N \sum _{i=1}^{max} \int\limits_{C(E)} a_i exp \left [ -c_i(E -
\epsilon_i)^2 \right ] E^n dE.
\label{7}\end{equation}
After some algebra we get the expressions for the moments as functions of the
parameters describing the height ($a_i$), the width ($c_i$) and the locations
of the maxima ($\epsilon_i$). In particular,
\begin{eqnarray}
M_1&=&N \sum_{i=1}^{max} \epsilon_i a_i \sqrt{\frac{\pi}{c_i}},\label{8}\\
M_2&=&N \sum_{i=1}^{max} a_i \sqrt{\frac{\pi}{c_i}} \left( \frac{1}{2c_i} +
\epsilon_i^2 \right),\label{9}\\
M_3&=&N \sum_{i=1}^{max} a_i \epsilon_i \sqrt{\frac{\pi}{c_i}} \left (
\frac{3}{2c_i} +
\epsilon_i^2 \right ),\label{10}\\
N&=& \left( \sum_{i=1}^{max} a_i \sqrt{\frac{\pi}{c_i}} \right) ^{-1}.
\label{11}\end{eqnarray}
According to the so called  {\it principle of moments}
\cite{brody,french,ivan} we expect that
if we identify the lower moments of two distributions, we bring these
distributions to approximate identity. In this paper we apply this
principle to the theory of molecular similarity. 
We assume that molecules have similar properties if their intensity
distributions and, consequently the corresponding moments, 
are approximately the same.

We propose that statistical moments of the intensity distributions can be
treated as a new kind of molecular descriptors. A very clear meaning has the
first moment, $M_1$, which describes the mean value of the distribution. In
a similar sense a colour index has been introduced in astronomy 
\cite{astr} - its value allows us to compare spectra of different stars 
(it carries an information about molecules forming the star). The second 
centered moment, $M_2^{\prime}$, is the variance which gives the width 
of the distribution. $M_3^{\prime\prime}$ is the skewness coefficient 
which describes the asymmetry of the spectrum. The kurtosis 
coefficient $M_4^{\prime\prime}$ is connected to the excess 
of the distribution. 

\section{Theory and the Model Spectra}

According to the method of moments, the shapes of two distributions are more
similar if the number of identical moments is larger. Similarity of
distributions in two- and three-moment approximations, in the context of the
construction of envelopes of electronic bands, has been analyzed in Refs.
\cite{biel1,biel2,biel3,biel4}. Analogously, we define similarity 
parameters ${\cal S}_k^{i_1i_2 \ldots i_k}$ ($k$ is the number of
properties taken into account in the process of comparison) as a normalized
information derived from a comparison of two distributions, referred to as
$\alpha$ and $\beta$:
\begin{eqnarray}
{\cal S}_1^{i_1}&=&\sqrt{D_{i_1}^2},\label{12}\\
{\cal S}_2^{i_1 i_2}&=&\sqrt{\frac{1}{2}\left ( D_{i_1}^2 + D_{i_2}^2
\right )} ,\;\;i_1<i_2\label{13}\\
\vdots\nonumber&\\
{\cal S}_k^{{i_1}{i_2}\ldots i_k}&=&\sqrt{\frac{1}{k} 
\left ( D_{i_1}^2 + D_{i_2}^2 + \ldots
D_{i_k}^2 \right )} ,\;\;\nonumber \\
& & {i_1}<{i_2}< \ldots i_k\label{14}\\
\vdots\nonumber &\\
{\cal S}_n^{i_1i_2\ldots i_n}&=&
\sqrt{\frac{1}{n}\sum\limits_{i=1}^n D_{i}^2}.
\label{15}\end{eqnarray}
Here $n$ is the total number of properties taken into account in the
comparison of the two spectra and $i_k=1,2, \ldots n\;\;(k=1,2,\ldots n)$,
correspond to a specific property. In particular, as the property number one
($i_k=1$) we take the first moment, as the property number two ($i_k=2$) we
take the second centered moment, number three ($i_k=3$) - the asymmetry
coefficient, number four ($i_k=4$) - the kurtosis coefficient. In this paper
we take $n=4$ and the corresponding similarity distances are defined as
follows:
\begin{equation}
D_1=1 - \exp \left [- \left ( M_1^{\alpha} - M_1^{\beta} \right )^2\right],
\label{16}\end{equation}
\begin{equation}
D_2=1-\exp\left[-\left(M_2^{\prime\alpha}-M_2^{\prime\beta}\right)^2
\right ],    
\label{17}\end{equation}
\begin{equation}
D_3=1-\exp\left[-\left(M_3^{\prime\prime\alpha}-M_3^{\prime\prime\beta}
\right)^2\right],
\label{18}
\end{equation}
\begin{equation}
D_4=1-\exp\left[-\left(M_4^{\prime\prime\alpha}-M_4^{\prime\prime\beta}
\right ) ^2 \right ].
\label{19}\end{equation}
The values of all the descriptors may vary from $0$ (identical properties) 
to $1$.

We also define an additional parameter which may be evaluated if both 
spectra we are going to compare are available:
\begin{equation}
\mathcal{D}=\frac{1}{2}
\int \limits_{C^{\prime}(E)}|I^{\prime\alpha}(E) - I^{\prime\beta}(E)| dE.
\label{20}
\end{equation}
This parameter is given by the integral of the module of the difference
between the compared distributions and is not related to the moments. In the
definition of $\mathcal{D}$, $I^{\prime}$ denotes the distributions
transformed so that their averages are the same. If we compare two
distributions of the same shape then $\mathcal{D}=0$. If two distributions
do not overlap at all, then $\mathcal{D}=1$. It is important to note that
the distribution moments are defined as numbers attached to a given spectrum
and the similarity distances $D_n$ are easily derived from the knowledge of
these numbers. The parameter $\mathcal{D}$, though it gives accurate
information about similarity of two spectra, is rather cumbersome since it
may be derived only if the complete spectra are given.

If two model molecules (or rather their spectra) are identical, up to the
accuracy determined by the considered properties, then all
${\cal S}_k^{i_1i_2 \ldots i_k}$ are equal to $0$. 
The maximum value of ${\cal S}_k^{i_1i_2 \ldots i_k}$ is $1$ and
corresponds to two spectra with no common features within the considered
set of properties.

The result of a comparison of two different objects depends not only on the
number of properties taken into account but also on their choice ($i_1$
or $i_2$ or $\ldots$  $i_n$).
Therefore the quantities ${\cal S}_k^{i_1i_2\ldots i_k}$ defined in Eq.
\ (\ref{12}) -\ (\ref{15}) should be averaged by 
taking all combinations of the indices $i_k$.
Thus, we define parameters $S_k$ as the appropriate averages of 
${\cal S}_k^{i_1i_2\ldots i_k}$:
\begin{equation}
S_k={n\choose k}^{-1}
\sum\limits_{i_1< i_2 < \ldots < i_k}^n {\cal S}_k^{i_1 i_2 \ldots
i_k}.
\label{21}
\end{equation}
In particular, in our case:
\begin{eqnarray}
S_1&=&\frac{1}{4}\sum\limits_{i_1=1}^4{\cal S}_1^{i_1},\label{22}\\
S_2&=&\frac{1}{6}\sum\limits_{i_1 < i_2}^4{\cal S}_2^{i_1i_2},\label{23}\\
S_3&=&\frac{1}{4}\sum\limits_{i_1 < i_2 < i_3}^4
{\cal S}_3^{i_1i_2i_3},\label{24}\\
S_4&=&{\cal S}_4^{i_1i_2i_3i_4}.
\label{25}
\end{eqnarray}

\section{Results and Discussion}

In order to illustrate our approach, we took model spectra consisting of two
bands, i.e. having two maxima ($max=2$):
\begin{equation}
I^{\gamma}(E) =  N\left[a_1\exp\left[-c_1(E-\epsilon_1)^2\right]+\right.
                  \left. a_2\exp\left[-c_2(E-\epsilon_2)^2\right]\right],
\label{26}
\end{equation}
where $\gamma=\{c_1,a_1,\epsilon_1,c_2,a_2,\epsilon_2\}$.
In order to see relations between molecular spectra, defined in Eq. 
\ (\ref{26}) and the similarity indices defined in Eqs. \ (\ref{16}) -\ 
(\ref{20}) and \ (\ref{22}) -\ (\ref{25}) in a simple and transparent way, we
study three sequences of spectra, where in each sequence only one parameter
has been modified: $c_2$ in sequence I, $a_2$ in sequence II, $\epsilon_2$
in sequence III. 
\begin{figure}
\includegraphics[width=8cm]{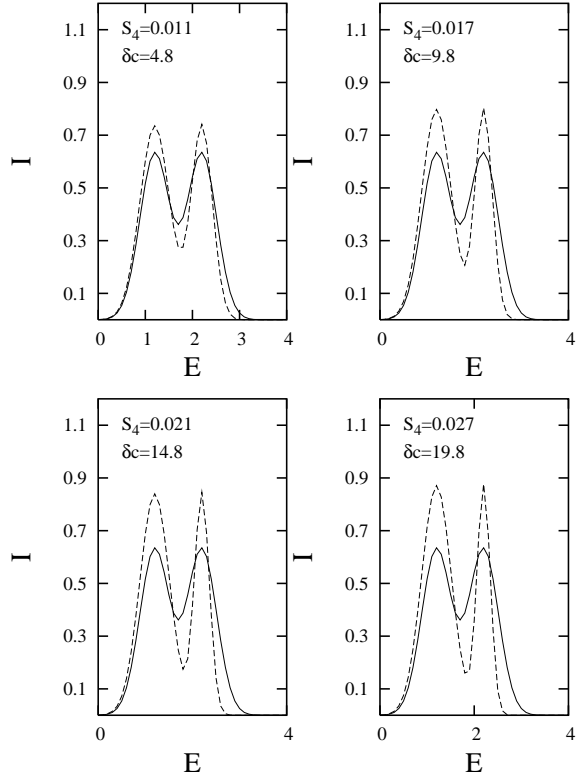}
\caption{ Two intensity distributions (solid and dashed lines) and
the corresponding similarity parameters $S_4$ (sequence I).}
\label{fig1}
\end{figure}
\begin{figure}
\includegraphics[width=8cm]{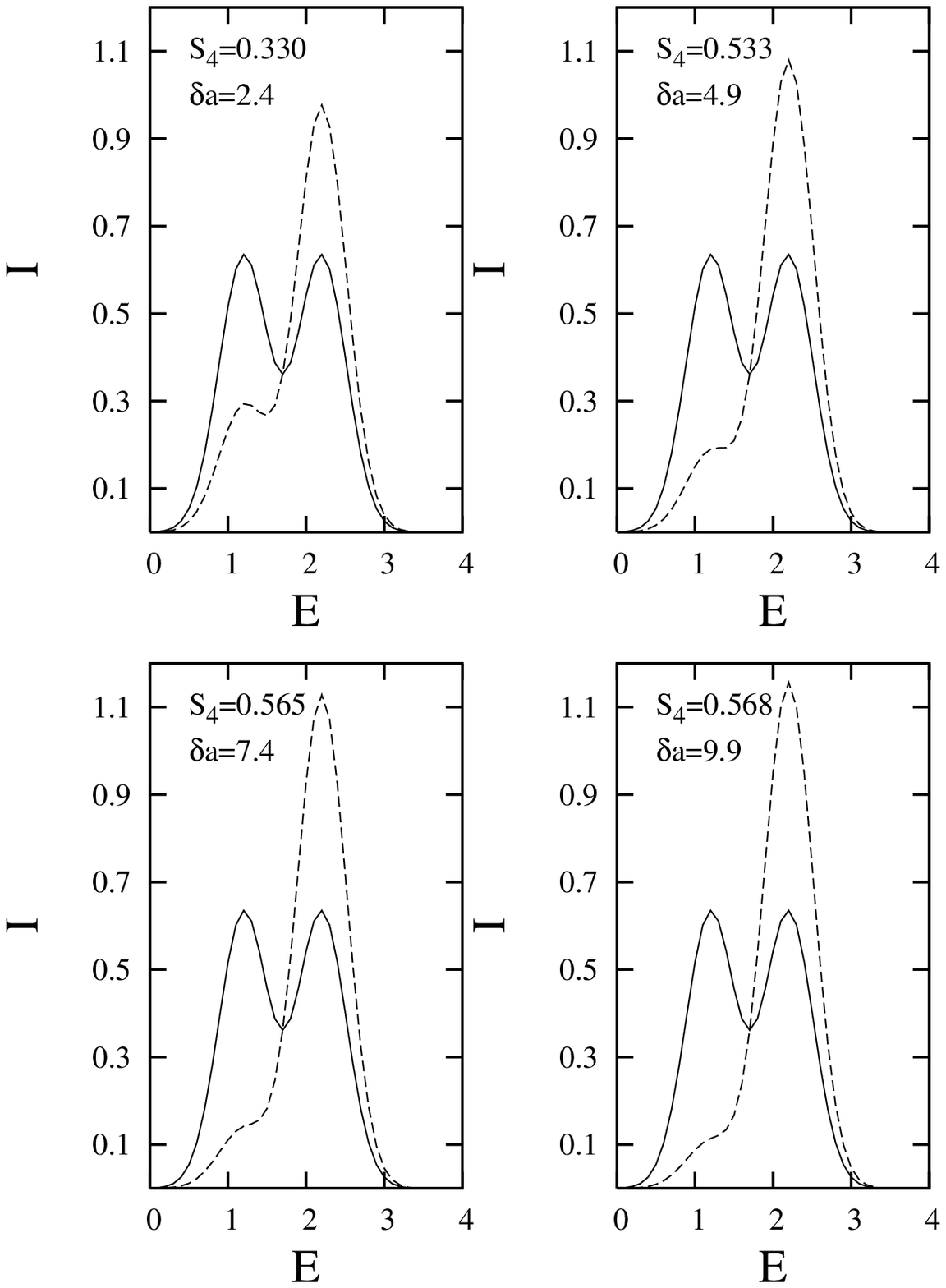}
\caption{ Two intensity distributions (solid and dashed lines) and
the corresponding similarity parameters $S_4$ (sequence II).}
\label{fig2}
\end{figure}
\begin{figure}
\includegraphics[width=8cm]{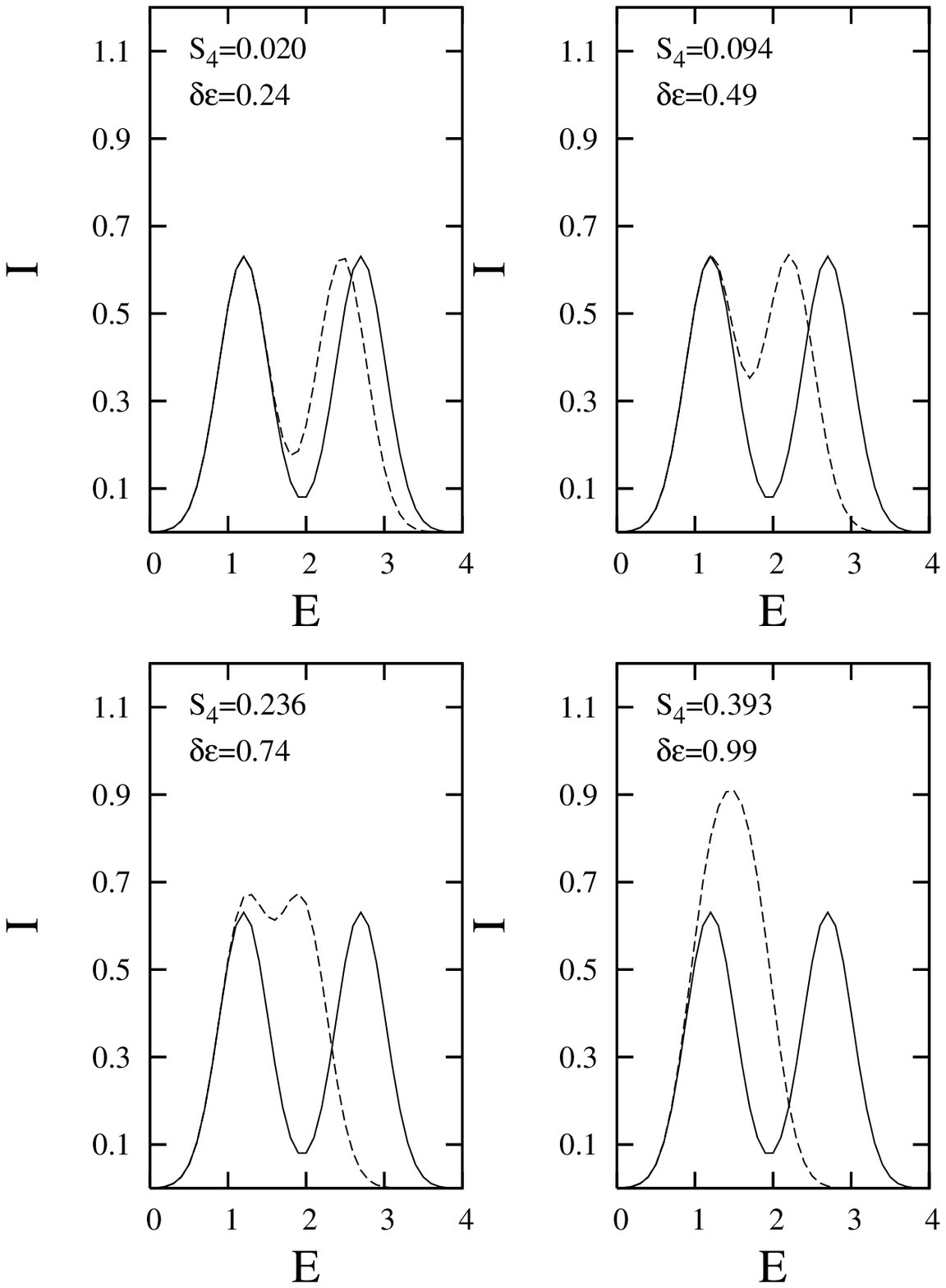}
\caption{ Two intensity distributions (solid and dashed lines) and
the corresponding similarity parameters $S_4$ (sequence III).}
\label{fig3}
\end{figure}
\begin{itemize}
\item[(a)]
Sequence I corresponds to the situation when a symmetric
spectrum consisting of two identical Gaussian distributions shifted relative
to each other by $\epsilon_2 - \epsilon_1=1$ ($a_1=a_2=1.0$,
$\epsilon_1=1.2$, $\epsilon_2=2.2$, $c_1=c_2=5.0$) transforms to a 
distribution in which the width of one of the Gaussians changes due to the
change of the parameter $c_2=5.0 + \delta c$, where $\delta c \in \langle
0;19.8 \rangle$. Then, we compare shapes of intensity distributions
$I^{\alpha}(E)$ and $I^{\beta}(E)$, where $\alpha=\{5.0,1.0,1.2,5.0,1.0,2.2\}$,
$\beta=\{5.0,1.0,1.2,5.0 + \delta c,1.0,2.2\}$.

In Fig. 1 spectra corresponding to $\delta c=0$ (solid lines) and $\delta c
> 0$ (dashed lines) are compared. In each case values of
$\delta c$ and
$S_4$ are given. A correlation between these two numbers and between shapes
of the spectra is clearly seen. The value of $S_4$ increases when the two
spectra become less similar.
\item[(b)]
Sequence II corresponds to the same symmetric spectrum as before
($a_1=a_2=1.0$, $\epsilon_1=1.2$, $\epsilon_2=2.2$, $c_1=c_2=5.0$)
transforming to the distributions in which the height of one of the
Gaussians changes due to the changes of $a_2=1.0 + \delta a$, where
$\delta a \in \langle 0; 9.9 \rangle$. Then, we compare shapes of intensity distributions
$I^{\alpha}(E)$ and $I^{\beta}(E)$, where $\alpha=\{5.0,1.0,1.2,5.0,1.0,2.2\}$,
$\beta=\{5.0,1.0,1.2,5.0,1.0 +\delta a,2.2\}$.

In Fig. 2 spectra corresponding to $\delta a=0$ (solid lines) and $\delta
a > 0$ (dashed lines) are compared.
In each case values of $\delta a$ and
$S_4$ are given.
The conclusions are similar to those in
the case of Fig. 1.
\item[(c)]
Sequence III corresponds to a similar situation as before,
except that the maxima in $I^{\alpha}$ are shifted by $1.5$ rather than by
$1$ ($a_1=a_2=1.0$,
$\epsilon_1=1.2$, $\epsilon_2=2.7$, $c_1=c_2=5.0$). $I^{\alpha}$
transforms to the distribution $I^{\beta}$ for which one of the gaussian
distribution changes the location of the second maximum
$\epsilon_2=2.7 - \delta \epsilon$, where
$\delta \epsilon \in \langle 0;0.99 \rangle$. 
Then, we compare shapes of intensity distributions
$I^{\alpha}(E)$ and $I^{\beta}(E)$, where $\alpha=\{5.0,1.0,1.2,5.0,1.0,2.7\}$,
$\beta=\{5.0,1.0,1.2,5.0,1.0,2.7 - \delta \epsilon\}$.

In Fig. 3 spectra corresponding to $\delta \epsilon=0$ (solid lines) and
$\delta\epsilon>0$ (dashed lines) are compared.  In each case
values of $\delta\epsilon$ and $S_4$ are given. The conclusions are similar
to those in the cases described by Figs. 1 and 2.
\end{itemize}
The molecular descriptors [statistical moments of $I^{\beta}(E)$] are
plotted in Fig. 4 versus $\delta c$ (sequence I), $\delta a$ (sequence II),
$\delta \epsilon$ (sequence III). In case of sequence I, it is clear that
the considered change of the spectrum leads to a decrease of the first
moment (the intensity is shifted towards smaller energies). The dispersion
of the whole distribution also decreases ($M_2^{\prime}$). The asymmetry of
the spectrum changes from totally symmetric ($M_3^{\prime\prime}=0$) to
asymmetric ($M_3^{\prime\prime}\ne 0$).  The kurtosis coefficient
$M_4^{\prime\prime}$ changes as it is presented, in a non-monotonic way.  It
is interesting that for $M_3^{\prime\prime}$ and $M_4^{\prime\prime}$ minima
appear for $\delta c \ne 0$. In the case of sequence II, with an increase of
$\delta a$ the first moment is shifted towards higher values and the
dispersion of the whole spectrum decreases. The asymmetry of the spectrum
decreases and the kurtosis parameter increases. In case of sequence III,
shifting the second maximum $\epsilon_2$ to the smaller energies results in
a distribution with one maximum instead of two and the intensity is shifted
towards smaller energies. In consequence the first moment decreases. The
whole distribution becomes more narrow and, consequently, we observe
decreasing of $M_2^{\prime}$. For all $\delta\epsilon$ distributions are
symmetric ($M_3^{\prime\prime}=0$) and the kurtosis parameter increases.
\begin{figure}
\includegraphics[width=10cm]{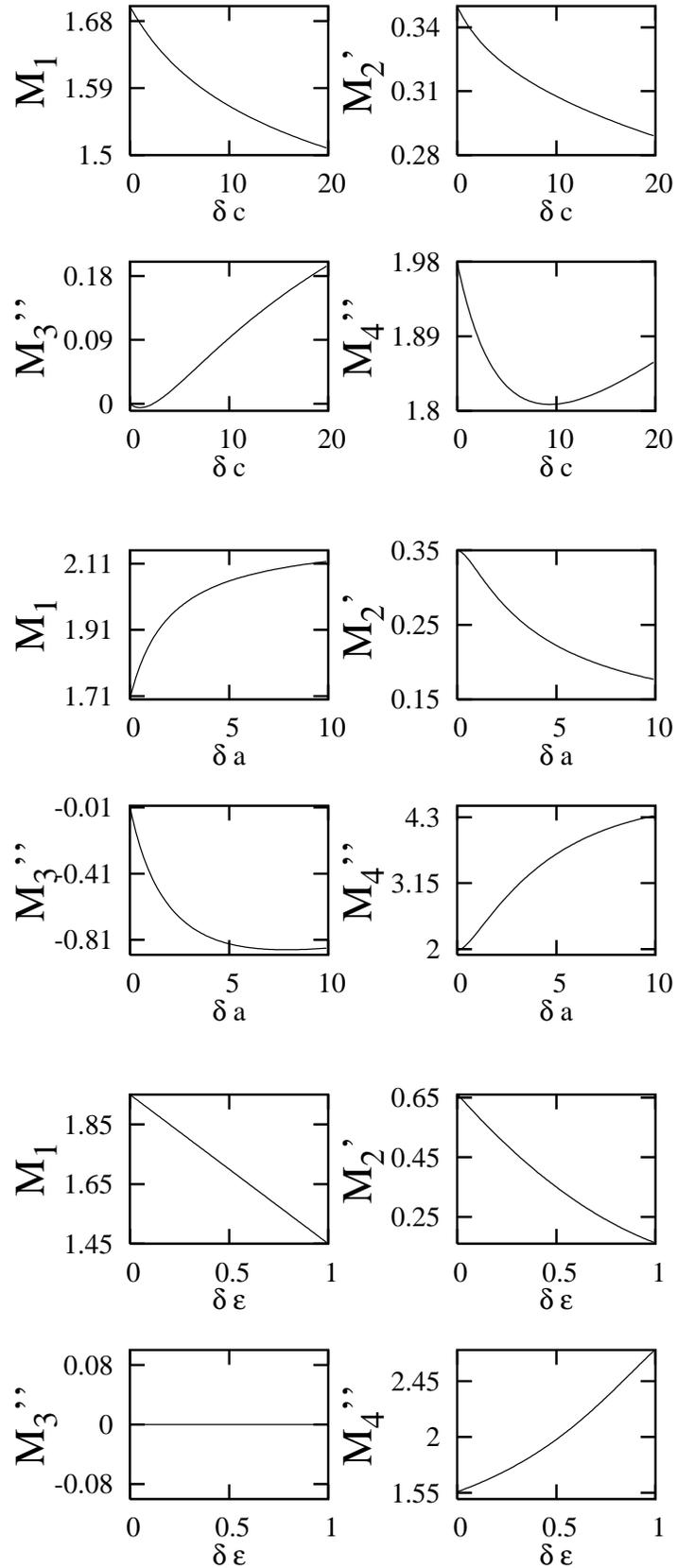}
\caption{Moments of the distributions as functions of $\delta c$ 
(sequence I), $\delta a$ (sequence II), $\delta \epsilon$ (sequence III).}
\label{fig4}
\end{figure}

Fig. 5 presents $D$ defined in Eqs. \ (\ref{16}) -\ (\ref{20}).
In the
case of sequence I, if $\delta c=0$, we compare two identical distributions
and all the descriptors are equal to zero. The most sensitive to the changes
of $\delta c$ is in this case $\mathcal{D}$, contrary to the other
descriptors which are nearly constant. The two distributions are rather
similar in sense of the average value, of the width, of the asymmetry and of
the kurtosis (the values of $D_1,D_2,D_3,D_4$ are small and the
corresponding curves cross). In case of sequence II, we observe small values
of $D_2$ and $D_1$, that indicates large similarity of the two distributions
in sense of the width and of the average values. For small values of $\delta
a$ we observe crossings between $D_3, D_4$ and $\mathcal{D}$. The most
sensitive to the changes of $\delta a$ is $D_4$. In case of sequnce III, the
behaviour of $D_1$ and $D_2$ is very similar. Both spectra are totally
symmetric ($M_3^{\prime\prime\alpha}=M_3^{\prime\prime\beta}=0$). Therefore
$D_3=0$ for all $\delta \epsilon$. $D_4$ and $\mathcal{D}$ cross and change
very substantially contrary to $D_1$ and $D_2$ which are nearly constant.
\begin{figure}
\includegraphics[width=5.5cm]{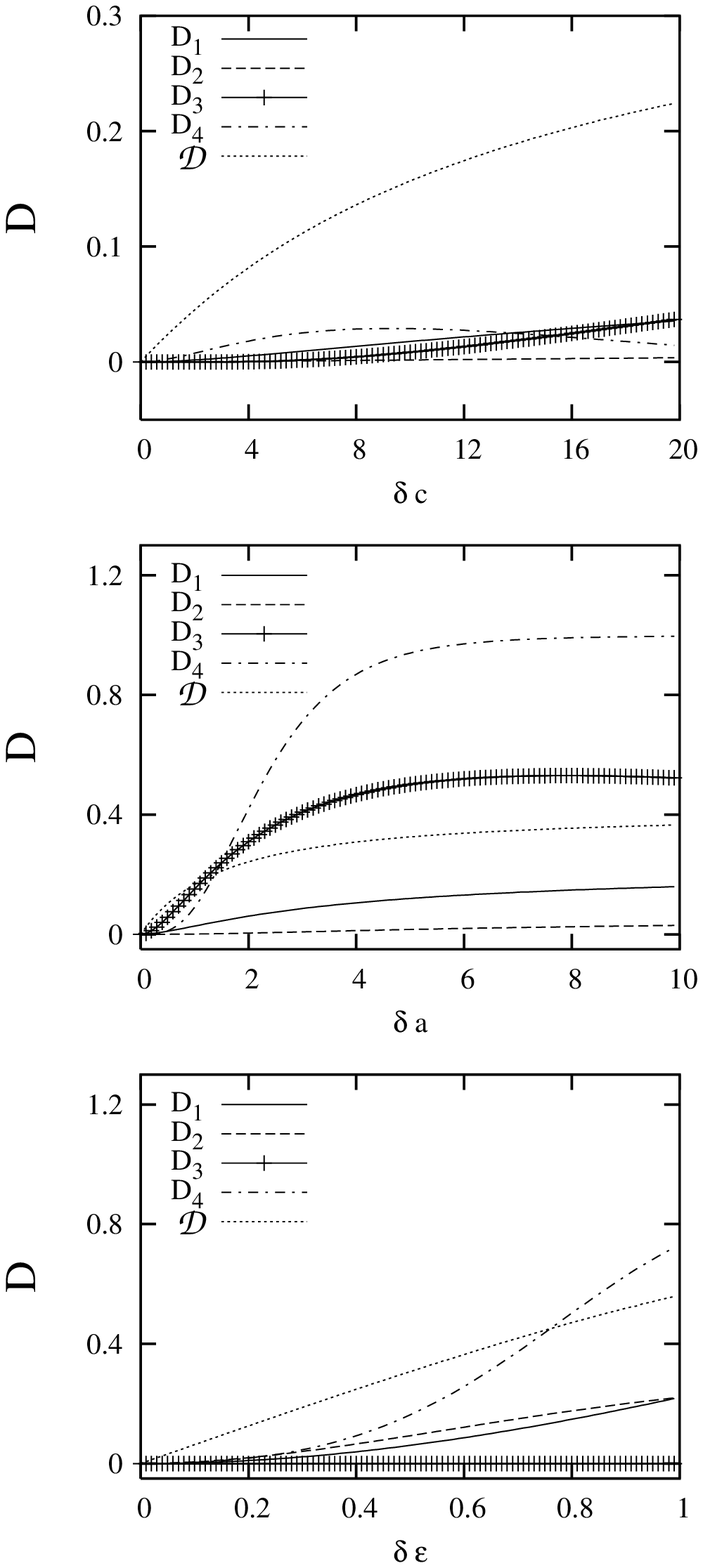}
\caption{ Parameters $D$ as functions of $\delta c$ (sequence I), 
$\delta a$ (sequence II), $\delta \epsilon$ (sequence III).}
\label{fig5}
\end{figure}
\begin{figure}
\includegraphics[width=5.5cm]{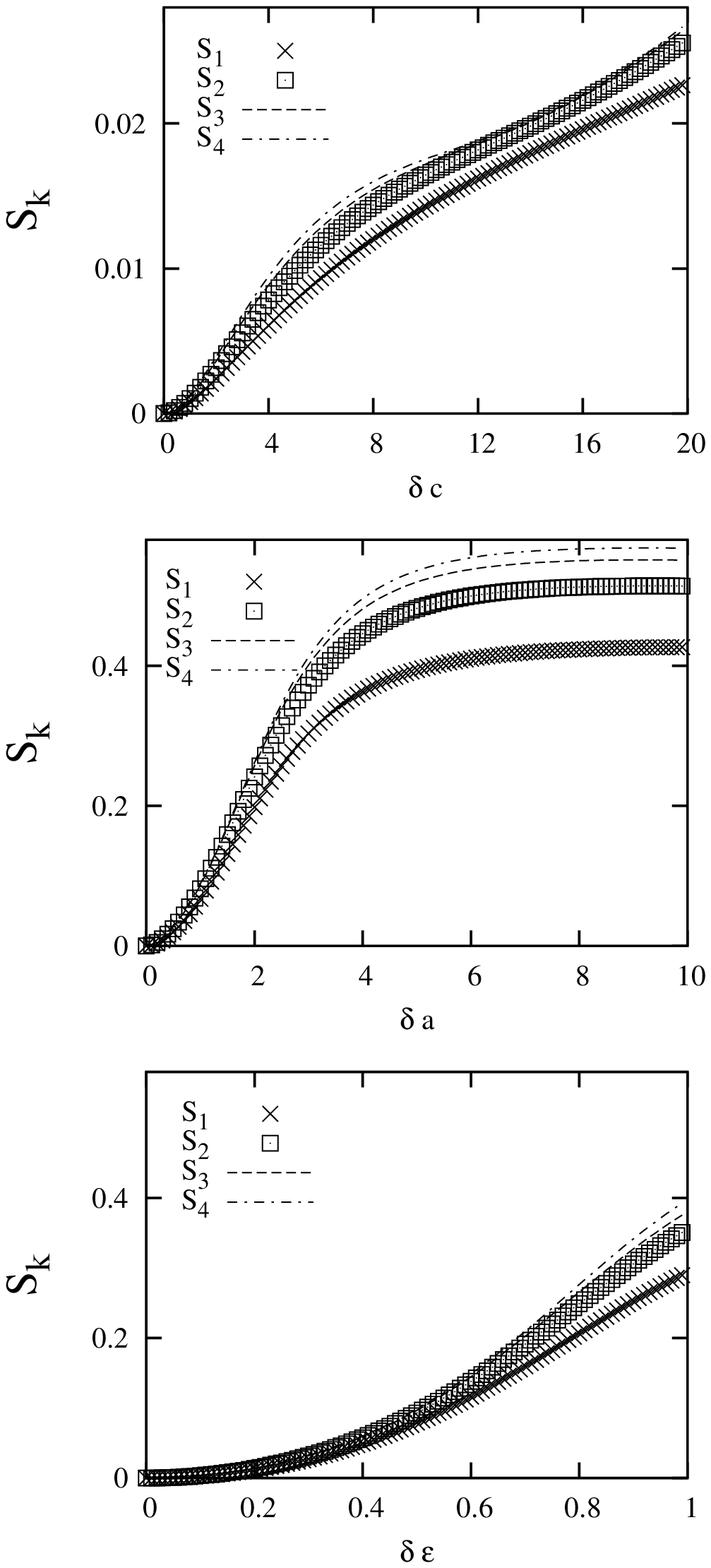}
\caption{ Parameters $S_k$ as functions of $\delta c$ (sequence I), 
$\delta a$ (sequence II), $\delta \epsilon$ (sequence III).}
\label{fig6}
\end{figure}

Fig. 6 presents similarity parameters $S_k$ for $k=1,2,3,4$ [Eqs. \
(\ref{22}) -\ (\ref{25})].  Small values of $S$ correspond to high similarity
of the model spectra. In particular, if $\delta c=0$ (sequence I) then
$S_k=0$ for all $k$. As we can see, $S$ is the smallest for $k=1$ and
increases with increasing $k$. Analogously to the sequence I,
$S_1<S_2<S_3<S_4$ for all $\delta a$ (sequence II) and for all $\delta
\epsilon$ (sequence III). Intuitively, we expect that two systems which are
similar to each other when only one property is considered may exhibit more
differences if we look at the systems in more detail, taking into account
more properties. These features can be seen in Fig. 6.

\section{Conclusions}

Statistical moments describe in an adequate way the degree of similarity of
two-band model spectra. Though the mathematical model describing shapes of
the spectra is relatively simple, it reflects the behaviour of real
molecular spectra. Three parameters: $c$, $a$ and $\epsilon$, influence
different aspects of the shapes of spectra and the resulting values of
$D$. In particular, parameters $D$ and corresponding $S$ are the
smallest if $a$ and $\epsilon$ are constant (sequence I). In these cases
spectra are only slightly modified by $\delta c$ (Fig. 1). Larger
differences of spectra are caused by parameter $\delta a$ , while $c$ and
$\epsilon$ are constant (sequence II). The influence of $\epsilon$ on
spectra is also large (sequence III). The additional parameter $\mathcal{D}$
introduces some independent information about spectra. Contrary to the case
of single-band model spectra studied in our previous paper \cite{biel5},
where its behaviour is very similar to $D_4$, here it apeears to be the most
sensitive index (sequence I).

Summarizing, we demonstrated that spectral density distribution moments can
be used for defining similarity indices of spectra. By grouping molecules
according to the spectral density distribution moments we can get a chance
to discover new characteristics in the field of molecular similarity and in
particular it may be a tool for studies in the area of computational
toxicology \cite{basak0,basak1,basak2}.
\acknowledgments
This work has been supported by Polish Ministry of Science and Information
Society Technologies, grant no 2 PO3B 033 25.


\begin{thebibliography}{33}
\bibitem{brody}
T. A. Brody, J. Flores, J. B. French, P. A. Mello, A. Pandey,
S. S. M. Wong, Rev. Mod. Phys. {\bf 53}, 385 (1981).
\bibitem{french}
J. B. French, V. K. Kota, Annual Review of Nuclear
and Particle Science, ed. J. D. Jackson, H. E. Gove, R. F.
Schwitters (Palo Alto, CA 1982) p. 35.
\bibitem{ivan}
V. S. Ivanov and V. B. Sovkov, Opt. Spectrosc. {\bf 74}, 30 (1993); 
Opt. Spectrosc. {\bf 74}, 52 (1993).
\bibitem{astr}
B. W. Carroll, D. A. Ostlie, An Introduction to Modern
Astrophysics, ed. Addison-Wesley Publishing Company Inc. (1996).
\bibitem{biel1}
D. Bieli\'nska-W\c{a}\.z, J. Karwowski,
   Phys. Rev. A {\bf 52}, 1067 (1995).
\bibitem{biel2}
D. Bieli\'nska-W\c{a}\.z, J. Karwowski,
   Advances in Quantum Chemistry {\bf 28}, 159 (1997).
\bibitem{biel3}
D. Bieli\'nska-W\c{a}\.z, J. Karwowski,
J. Quant. Spec. Rad. Transfer {\bf 59}, 39 (1998).
\bibitem{biel4} 
D. Bieli\'nska-W\c{a}\.z, in
Symmetry and Structural Properties of Condensed Matter, World
Scientific Singapore 1999, pp. 212-221.
\bibitem{biel5}
D. Bieli\'nska-W\c{a}\.z,P. W\c{a}\.z, S. C. Basak, R. Natarajan,
{\it Statistical Theory of Spectra as a Tool in Molecular Similarity},
submitted for publication.
\bibitem{basak0}
S. C. Basak, B. D. Grunwald G. E. Host, G. J. Niemi, and S. P. Bradbury,
Environ Toxicol. Chem. {\bf 17}, 1056 (1998).
\bibitem{basak1}
S. C. Basak, K. Balasubramanian, B. D. Gute and D. Mills, 
J. Chem. Inf. Comput. Sci. {\bf 43}, 1103 (2003).
\bibitem{basak2}
S. C. Basak, B. D. Gute, D. Mills, and D. Hawkins, J. Mol. Struct.
(Theochem) {\bf 622}, 127 (2003).
\end{thebibliography}
\end{document}